\documentclass[twocolumn]{elsart5p}

\usepackage{graphicx,amssymb,color}
\journal {NIM A}

\begin{document}

\begin{frontmatter}



\title{Replika Mirrors - Nearly Loss-Free Guides For Ultracold Neutrons}


\author[ILL]{C. Plonka\corauthref{cor}},
\corauth[cor]{tel: +33 4\,7620\,7629; fax: +33 4\,7620\,7777}
\ead{plonka@ill.fr}
\author[ILL]{P. Geltenbort},
\author[ILL]{T. Soldner},
\author[SDH]{H. H{\"a}se\thanksref{now}}

\thanks[now]{www.s--dh.de}

\address[ILL]{Institut Laue Langevin, 6 rue Jules Horowitz,
38042 Grenoble, France}
\address[SDH]{S--DH GmbH, Hans--Bunte--Stra\ss e 8--10, 69123 Heidelberg, Germany}

\begin{abstract}
The reflectivity of ultracold neutron (UCN) guides 
produced with a dedicated technique called {\sc Replika} has been studied. The
guides are made of nickel, where the surface quality was copied
from a glass layer. This results in a surface 
roughness smaller than 10~\AA. The reflectivity 
was measured to be 99.9~\% or higher. 
Those guides can be used at present or future UCN 
sources to transport UCN over long distances to the respective
experiments without significant losses.

\end{abstract}

\end{frontmatter}

\section{Introduction}
\label{sec:1}
Ultracold neutrons (UCN) have velocities below 10~m/s and wavelengths
around 1000~\AA. They are totally reflected under each angle 
of incidence from certain materials and hence storable for hundreds of 
seconds in volumina made of those materials. 
This remarkable feature has put UCN in a favorite role to investigate 
fundamental properties of the neutron like its lifetime 
\cite{Serebrov,Arzumanov}, its electric dipole moment 
\cite{Baker,PNPI}, or its quantum states in the Earth's gravitational field 
\cite{Nesvizhevsky2}.\\
UCN are guided in reflective tubes
from the source to the experiment. The roughness of the guide surface
is crucial for its transport properties. On rough surfaces, neutrons are
non-specular reflected, resulting in a diffusive transport. Therefore
the number of reflections for a neutron to travel a given distance decreases
compared to a smooth surface with 
specular reflection and the transmission of the guide will be reduced 
significantly.
To overcome this loss, essential R\&D work on UCN guide properties is carried 
out at many laboratories \cite{Nesvizhevsky1,Brys, Atchison,Altarev, 
Kawabata, Kawabata2, Brys2}.\\
The UCN guides we report on are made of a metallic surface with a high
{\sc Fermi} potential - natural nickel - whose 
high neutron-optical quality is achieved in a replication
process first used for UCN application some 20 years ago 
\cite{Steyerl}. The technique
has been revisited by the Sputter-D{\"u}nnschicht company in Heidelberg
(S--DH).

\section{Interaction of UCN with material surfaces -- UCN transport}
\label{sec:2}
The coherent interaction of neutrons with a surface is described 
by the {\sc Fermi}-Potential $V_f=V+iW$,
where $V$ and $W$ depend on the nuclear properties of the surface atoms 
seen by the neutrons \cite{Golub}. The real part of the potential $V$ sets
a critical velocity $v_c = \sqrt{2V/m}$, with the neutron
mass $m$. Neutrons with a normal velocity component 
$v_{\scriptscriptstyle{\perp}}<v_c$ are totally reflected from the 
surface with a probability given by the reflectivity coefficient 
\mbox{$R = 1-2\,\frac{W}{V} 
\left ( \frac{v^2_{\scriptscriptstyle{\perp}} }
{2\,V/m-v^2_{\scriptscriptstyle{\perp}} } \right )^{1/2}$.}
\begin{figure*}[t]
\centering{
\includegraphics[width=.7\textwidth]{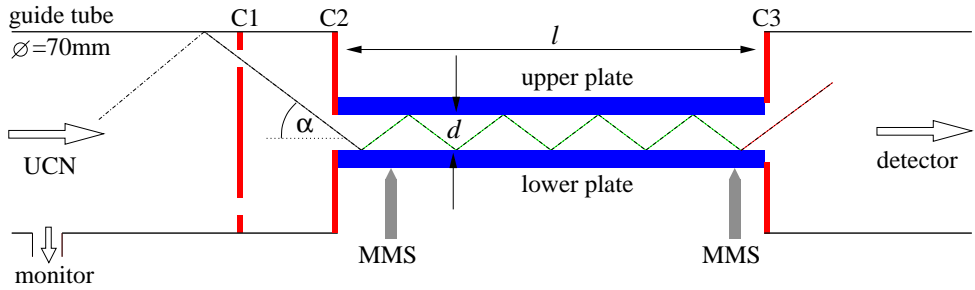}
\caption{\label{fig:Setup}Experimental set-up (not to scale). 
UCN from
the {\sc Steyerl} turbine at ILL pass the collimation C1 (two symmetrical openings up and 
down) and the transport channel between two parallel plates and leave towards the 
detector. The adjusted distance as shown in the draft would correspond 
to $d=2~\rm mm$. For larger distances, the UCN see the same opening angle 
into the channel and do not collide with the lower plate. The whole set-up
was placed in vacuum at a pressure below $5\times10^{-4}\rm mbar$.}
}
\end{figure*}
The dimensionless ratio \mbox{$\eta=W/V$} is called the energy-independent 
wall-loss probability
per bounce. It describes the UCN losses on the surface 
material due to absorption and inelastic scattering.\\
Nickel is commonly used as guide material due to its 
high {\sc Fermi}-Potential of $V$=252~neV, which corresponds to a 
critical velocity \mbox{$v_{\rm c}=7~\rm {m/s}$}.
For UCN with velocities close to the critical velocity $v_c$,
the reflectivity decreases rapidly. 
However, sufficiently below $v_c$ the reflectivity remains higher than 99.9~\%.\\
Such a high reflectivity would allow nearly loss-free UCN transportation
if only some hundreds of reflections over
typical distances of meters to the experiment are needed as it is the case
for specular transport.
This is no longer true, however, if non-specular reflections come into play.
In first order, the probability for
diffusive reflection of a wave on a rough surface is given by 
$(\Delta d/\lambda)^2$, where $\Delta d$ is the average surface roughness
and $\lambda$ the normal wavelength component. Therefore, for low loss
transportation of UCN, the surface
roughness has to be smaller than 50~\AA~to avoid
diffusive loss rates exceeding those from absorption and inelastic scattering.
\section{Replika guides}
\label{sec:3}
At the UCN facility PF2 \cite{Steyerl} of the Institut Laue Langevin (ILL), 
very cold neutrons \mbox{($v<100~\rm m/s$)} are extracted vertically from the 
liquid D$_2$ source  
of the high-flux reactor towards the so-called {\sc Steyerl}-turbine, 
where they are Doppler-shifted 
to the UCN regime. The cryogenic vertical guide, the curved feeding guide,
the turbine blades and the following exit guides are
made of nickel on an all-metal basis. A detailed description is given in \cite{Steyerl}.
The main idea of this so-called {\sc Replika} technique is to copy the
surface quality of a glass substrate to the guide surface.
These metal guides offer high mechanical flexibility and stability and 
withstand the high 
radiation level close to the reactor core.\\
The revisited {\sc Replika} plates were manufactured at \mbox{S--DH} in the following way: 
2000~\AA~nickel was sputtered onto Borofloat glass, gently solved, 
and then glued up-side-down onto an Aluminum plate as mechanical support.
The dimensions of the plates are $500\times70\times8~{\rm mm^3}$.
The roughness was measured using x-rays to 
be less than 10~\AA~over the whole surface.
\section{Measurement of the reflectivity}
\label{sec:4}
To measure the neutron reflectivity of such {\sc Replika} plates, 
two of them are installed as a neutron guide of length
$l=500~\rm mm$ with open sides, as shown in fig.~\ref{fig:Setup}.
Two collimators C1 and C2, made of cadmium, 
define a mean angle $\alpha=(40\pm6)^{\circ}$, 
under which UCN can enter the guide. 
The upper plate is fixed, while the lower one is vertically movable. 
The distance $d$ between the plates is adjustable
by means of micrometer-screws (MMS) with a reproducibility 
$\delta d<0.05\rm mm$. UCN have to undergo many specular reflections
before being counted in the detector. 
Those UCN, which leave the horizontal parallel plate system to its open sides, 
are immediately absorbed to avoid reentering.
A monitor in front of the guide monitors 
the flux of the incoming UCN.
Gaseous detectors with low $^3$He admixtures are used.
Count rates were around 0.3~cps and 1~cps 
for the detector and the monitor, respectively.\\
The average UCN velocity at the TES position of the PF2 facility
is $\sim\,$8~m/s.\\
The background of the detector was measured by a) putting
an UCN absorber between the plates and by b) closing the
diaphragm C3 towards the detector completely. Both measurements yield 
a similar background rate around $8\times 10^{-3}~\rm cps$. This implies that the 
background mainly arises from thermal reactor neutrons which can 
penetrate the detector shielding. The background rate of the monitor 
was found to be negligible.\\
\section{Analysis}
\label{sec:5}
\begin{figure}[b]
\includegraphics[width=.5\textwidth]{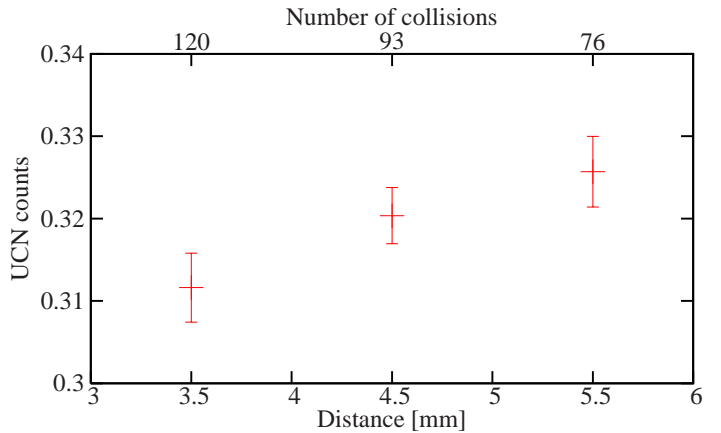}
\caption{\label{fig:Data}Normalized counts $T$ as a function of the distance
between the plates. The indicated number of collisions are calculated for the mean angle
of incidence $\alpha=40^{\circ}.$}
\end{figure}
Fig.~\ref{fig:Data} shows the counts $T$
of the detector after background correction and normalization 
to the monitor data. $T$ decreases with smaller distances between 
the plates as expected: For smaller 
distances the number of specular reflections 
increases, according to $n(\alpha)=l/d\,\tan \alpha$,
while the detection probability decreases with $T\propto R^n$.\\
The ratio of two measured $T$ values at different distances $d$ can be 
expressed by:
\begin{equation}
\label{eqn:CalculationR}
\quad\quad\quad\quad \frac{T_1}{T_2}=
\frac{\langle R^{n_1(\alpha)}\rangle_{\alpha}}
     {\langle R^{n_2(\alpha)}\rangle_{\alpha}} 
=
\frac{\langle R^{l/d_1\tan \alpha}\rangle_{\alpha}}
     {\langle R^{l/d_2\tan \alpha}\rangle_{\alpha}}.
\end{equation}
The term on the right hand side is calculated 
numerically as a function of the reflectivity $R$
and is compared to the 
measured count ratios. An isotropic UCN distribution between
$\alpha_{\rm min}$ and $\alpha_{\rm max}$ is assumed. It was shown, that
the results do not depend strongly on the form of the input distribution
as only the difference in the number of reflections enters.\\
By evaluating the different combinations of the measured distances $d$
we obtain a reflectivity of at least 99.9(1)\,\%.
The loss probability per bounce therefore is smaller than $10^{-3}$.
\section{Conclusion}
\label{sec:6}
All-metal guides produced by a replication process
have shown their potential during more than 20 years of use at the UCN/VCN 
facility PF2 of the ILL. They offer high optical performance for both VCN and UCN, 
are resistant against high radiation and due to their thinness they are of high
mechanical flexibility.\\
The revisited {\sc Replika} technique yields guides with a surface
roughness $<10$~\AA. Their reflectivity for UCN under
realistic experimental conditions was measured to be at least 99.9(1)~\% per
bounce, close to the theoretical expectations. With this technique,
UCN guides of several meters length for new UCN sources, but also upgrades of 
existing guide systems, can be produced.


\section{Acknowledgement}
\label{sec:7}
We are grateful to V.~Nesvizhevsky, S.~Petoukhov and I.~Altarev
for helpful discussions.

\end{document}